\documentclass[11pt,twoside]{article}
\pdfoutput=1
\usepackage{asp2010}

\resetcounters

\begin{document}

\title{The CCAT Software System}
\author{Tim~Jenness$^1$, Adam~Brazier$^1$, Kevin~Edwards$^2$,
  Gaelen~Marsden$^3$, Thomas~Nikola$^1$, Volker~Ossenkopf$^4$,
  Steve~Padin$^5$, Jack~Sayers$^5$, Reinhold~Schaaf$^6$, and Martin~Shepherd$^5$
\affil{$^1$Department of Astronomy, Cornell University, Ithaca NY,
  14853, USA}
\affil{$^2$Department of Physics, University of Waterloo, Waterloo, ON
N2L~3G1, Canada}
\affil{$^3$Department of Physics and Astronomy, University of British
  Columbia, 6224~Agricultural~Road, Vancouver, BC V6T~1Z1, Canada}
\affil{$^4$KOSMA, I.\ Physikalisches Institut, Universit\"{a}t zu
  K\"{o}ln, Z\"{u}lpicher Str.\ 77, 50937 K\"{o}ln, Germany}
\affil{$^5$California Institute of Technology, 1200 E California Blvd,
  Pasadena, CA 91125, USA}
\affil{$^6$Argelander-Institut f\"{u}r Astronomie, Universit\"{a}t
  Bonn, Auf dem H\"{u}gel 71, 53121 Bonn, Germany}}

\begin{abstract}
  CCAT will be a 25-meter telescope for submillimeter astronomy
  located at 5600\,m altitude on Cerro Chajnantor in northern Chile.
  CCAT will combine high sensitivity, a wide field of view, and a
  broad wavelength range (0.35 to 2.1\,mm) to provide an unprecedented
  capability for deep, large-area multicolor submillimeter surveys. It
  is planned to have a suite of instruments including large format KID
  cameras, a large heterodyne array and a KID-based
  direct detection multi-object spectrometer. The remote location
  drives a desire for fully autonomous observing coupled with data
  reduction pipelines and fast feedback to principal investigators.
\end{abstract}

\section{Key Drivers for Software Design}

CCAT \citep{2012SPIE.8444E..2MW,2013AAS...22115006G} has two key
drivers for the software design:

\textbf{Instrument Data Rates:} Next generation sub-millimeter
cameras with 40,000 to 65,000 detectors and reading out at 3 kHz can
generate data at a peak rate of $\sim$ 7 Gbps and generate 30 to 60 TB per
day.

\textbf{Remote Location:} The Observatory is located on Cerro Chajnantor above
the ALMA plateau at an altitude of 5600\,m. Given the remoteness
of the site, the Observatory is designed to operate autonomously to
the extent possible and support remote control from the base facility
in San Pedro de Atacama and/or partner sites.

\section{Proposed Instrumentation}

The CCAT project has identified four major instruments to achieve its
science goals. SWCam \citep{2013AAS...22115007S} is a 50,000-70,000
detector camera operating in the 350\,\micron\ and 450\,\micron\
windows with a short-wavelength goal of 200\,\micron. LWCam \citep{2013AAS...22115008G}
is a 40,000 detector camera operating in 5-6 bands between
750\,\micron\ and 2.1\,mm with a long-wavelength goal of 3.3\,mm. X-Spec
\citep{2013AAS...22115009B} is a multi-object spectrometer with
$\sim$\,100 beams on the sky, each covering a frequency range of
190-520\,GHz in two bands simultaneously with spectral resolutions
between 400 and 700. CHAI \citep{GoldsmithCHAI2012} is a
2$\times$64-element heterodyne array operating in the 600\,\micron\
and 350\,\micron\ bands.

\section{Software Systems}

\begin{figure}
\includegraphics[width=\textwidth]{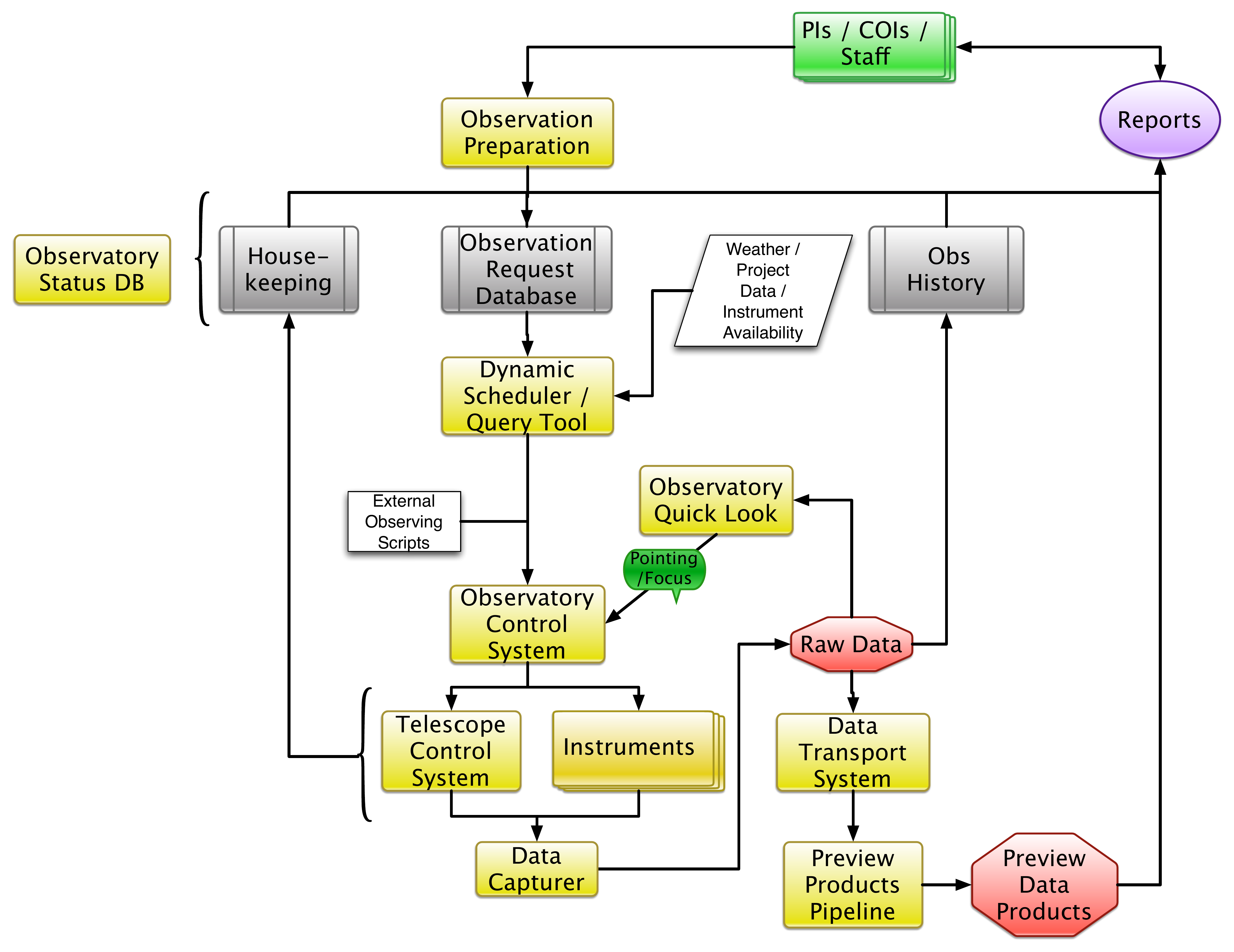}
\caption{Observations are prepared and stored by the OT and queried by
  the scheduler. The OCS executes observations and the Data Capturer
  collates the resulting data. The files are then transferred to the
  base facility and preview products are generated for review by the PI.}
\label{fig:P10_overview}
\end{figure}

An outline of the CCAT software and how the systems are connected is
given in Fig.\ \ref{fig:P10_overview}. This section provides more
detail on indvidual components.

\textbf{Observation Preparation:} CCAT is exploring the feasibility
of building onto existing submillimeter observation preparation tools (OTs)
such as the ALMA-OT \citep{2012SPIE.8451E..1AB}, the JCMT-OT
\citep{2002ASPC..281..453F} and Herschel-Spot
\citep{2009ESAB139.14R,2007AAS...210.1106F}. OT support of CCAT's
planned multi-object spectrograph instrument will require the addition
of new capabilities to whichever OT solution is chosen.

\textbf{Dynamic Scheduler:} CCAT is a flexibly-scheduled telescope and
will use a dynamic scheduler to determine what should be observed
next, similar to those employed at other sub-millimeter telescopes
such as JCMT \citep{2002ASPC..281..488E,2004SPIE.5493...24A} and ALMA
\citep{2007ASPC..376..673L}. The Scheduler will have a manual mode
allowing an observer to query the system but the aim is for the
scheduler to continually monitor the telescope environment and
observing queue and ensure that the best Minimum Schedulable Block (MSB) \citep[see
e.g.][for a more detailed description of an
MSB]{2011tfa..confE..42J} for the current
conditions is always observed and the telescope is never idle. One
complication in the sub-millimeter is the paucity of flux calibrators
in the sub-millimeter sky. This sometimes requires that the flux
calibrator is observed some time after the observation that requires
it. The scheduler must keep track of the calibrations requirements and
schedule them when appropriate.

\textbf{Observation Execution:} The Observatory Control System (OCS)
provides command and control and monitoring of all observatory systems
relating to observing. It uses a lightweight scripting execution
engine with guaranteed latencies. The OCS is influenced by experience
from earlier telescopes such as CBI, QUIET and the OVRO 40m \citep[see
e.g.][]{2013ApJ...768....9B}.

\textbf{Data Capturer:} The Data Capturer (DC) takes the output from the
instruments, telescope control system and other systems and collates
them into output data files. The large data rates, and the need to
support simultaneous acquisition by multiple instruments, led us to a
distributed approach where each subsystem writes its own data files to
disk and the DC writes a small header file with hierarchical linkage
to the individual files.

\textbf{Data Transport:} There will only be limited disk space at the summit
and data will immediately be transferred to the Base Facility via a
fiber connection.

\textbf{Preview Products Pipeline:} Raw data will be shipped to the
CCAT Data Archive on disks as network bandwidth is prohibitive. To
provide data quality assessment a pipeline will run at the Base
Facility to generate usable products suitable for remote observation
assessment.

\textbf{Data Reduction Challenges:} Reducing data from continuum
cameras with data rates 100 times larger than current instruments,
such as SCUBA-2 on JCMT \citep{2013MNRAS.430.2513H}, will be
challenging. Some thoughts on this can be found in
\citet{P14_adassxxiii}. Similarly CHAI has a data rate 64 times larger
than the HARP instrument on JCMT \citep{2009MNRAS.399.1026B}.

\textbf{Observation Management:} The remote autonomous operation
requires that there is a robust system in place for tracking Minimum
Schedulable Blocks and associated observations. Time accounting will
be automated by default with the ability to reject observations either
automatically, via QA assessment, or from human intervention. PIs will
be able to track the status of their project at any time and modify
their Science Program based on feedback from observations already
taken.

\textbf{Observatory Status:} All information generated by the observatory will
be logged permanently in an Observatory Status Database. The OSD will
store instrument house keeping data (possibly 0.5\,TB/day), observation
management information, calibration history and other digital data.

\textbf{Survey Pipeline:} Not explicitly listed in Fig.\ 1, the
processing of survey data will be coordinated through the CCAT Data
Archive (CDA). Reduced data products will be archived permanently and
will be made available via VO protocols once the proprietary period
expires. The CDA will also endeavor to reduce PI data if standard
observing modes were used.

\acknowledgements The CCAT Engineering Design Phase is partially
supported by funding from the National Science Foundation via AST-1118243.

\end{document}